\documentclass[showpacs,preprintnumbers,amsmath,amssymb]{revtex4}

\usepackage{graphics}
\usepackage{bm}

\begin{document}

\title{Regular and Black Hole Skyrmions with Axisymmetry}

\author{Nobuyuki Sawado and Noriko Shiiki}
\affiliation{Department of Physics, Tokyo University of Science, Noda, 
Chiba 278-8510, Japan}
\author{Kei-ichi Maeda$^{1,2,3}$ and Takashi Torii$^2$}
\affiliation{$^1$Department of Physics, Waseda University, Shinjuku, Tokyo 169-8555, Japan \\
$^2$Advanced Research Institute for Science and Engineering, Waseda University, 
Tokyo 169-8555, Japan \\
$^3$Waseda Institute for Astrophysics, Waseda University, Shinjuku, Tokyo 169-8555, Japan}
\date{\today}


\begin{abstract} 

It has been known that a $B=2$ skyrmion is axially symmetric. 
We consider the Skyrme model coupled to gravity and obtain static 
axially symmetric regular and black hole solutions numerically. 
Computing the energy density of the skyrmion, we discuss the effect of gravity 
to the energy density and baryon density of the skyrmion.
\end{abstract}

\pacs{04.70.Bw, 12.39.Dc, 21.60.-n \\
{\it Keywords}: Black holes, Skyrmions, Deuterons}

\maketitle


\section{Introduction}

The Skyrme model is an effective theory of QCD constructed by pion fields alone. 
Baryons are incorporated as topological solitons in this model, called skyrmions,  
and the topological charge corresponds to the baryon number 
$B$~\cite{Shiiki:skyrme,Shiiki:witten}. 
Skyrme introduced a hedgehog ansatz for the pion fields 
to obtain $B=1$ spherically symmetric soliton solutions.
The detailed analysis for the property of the $B=1$ skyrmion as a nucleon 
was performed in Ref.~\cite{Shiiki:adkins&nappi&witten} upon quantisation of  
the collective coordinate. Later Braaten and Carson obtained $B=2$ skyrmions 
numerically and showed that $B=2$ skyrmions are axially symmetric~\cite{Shiiki:braaten&carson}. 
Interestingly, multi-skyrmions with $B > 2$ exhibit various discrete symmetries analogously to  
multi-BPS monopoles~\cite{Shiiki:houghton&manton&satcliffe}.   

The Einstein-Skyrme system in which Skyrme fields couple with gravity was 
first considered by Luckock and Moss~\cite{Shiiki:luckock&moss} where  
the Schwarzschild black hole with Skyrme hair was obtained numerically. 
This is a counter example of the no-hair conjecture for black holes~\cite{Shiiki:ruffini&wheeler}.  
They observed that the presence of the horizon in the core of skyrmion 
unwinds the skyrmion, leaving fractional baryon charge outside the horizon. 
The full Einstein-Skyrme system was solved later to obtain spherically symmetric 
black holes with Skyrme hair~\cite{Shiiki:droz&huesler&straumann} and regular gravitating 
skyrmions~\cite{Shiiki:bizon&chmaj}. 

In this paper we shall study regular and black hole solutions 
of $B=2$ skyrmions with axisymmetry. 
The recent study by Hartmann, Kleihaus and Kunz showed the Einstein-Yang-Mills-Higgs 
theory possess axially symmetric monopole and black hole solutions~\cite{Shiiki:kleihaus&kunz,
Shiiki:hartmann&kleihaus&kunz}. 
Therefore, we shall follow their numerical technique to solve the Einstein-Skyrme model 
with axial symmetry. 
Computing the energy density and charge density, we shall show how gravity 
affects to the skyrmion. 

\section{The Einstein-Skyrme System}
 
After a long history, the Skyrme model was proved to be an effective theory of 
QCD by E. Witten in Ref.~\cite{Shiiki:witten}. 
At low energy the $SU(2)\times SU(2)$ chiral 
symmetry of the strong interaction is broken spontaneously and the 
Skyrme Lagrangian respects this symmetry. 
The Skyrme model coupled to gravity is defined by the Lagrangian,     
\begin{eqnarray*}
	{\cal L}={\cal L}_S + {\cal L}_G
\end{eqnarray*}
where
\begin{eqnarray}
	{\cal L}_S &=& \frac{f_\pi^2}{16}g^{\mu\nu}{\rm tr}\left(U^{-1}
	\partial_{\mu}UU^{-1}\partial_{\nu}U\right) \nonumber \\
	&& + \frac{1}{32a^2}g^{\mu\rho}g^{\nu\sigma}{\rm tr}\left(\left[U^{-1}\partial_{\mu}
	U, U^{-1}\partial_{\nu}U\right]\left[U^{-1}\partial_{\rho}U, U^{-1}
	\partial_{\sigma}U\right]\right) \label{ls} \\
	{\cal L}_G &=& \frac{1}{16\pi G}R  \label{lg}
\end{eqnarray} 
and $U$ is an $SU(2)$ valued scalar field of pions. 
$f_{\pi}\sim 186$ MeV is the pion decay constant. 

To solve the field equations, ansatz for the metric and Skyrme fields are required. 
Let us introduce an ansatz for the metric given in Ref.~\cite{Shiiki:kleihaus&kunz}
\begin{eqnarray}
	ds^2 = -fdt^2 + \frac{m}{f}(dr^2 + r^2 d\theta^2) + \frac{l}{f} 
	r^2 \sin^2\theta d\varphi^2
\end{eqnarray}
where $f=f(r,\theta), \,\, m=m(r,\theta), \,\,{\rm and}\,\, l=l(r,\theta)$. 

The axially symmetric Skyrme fields can be parameterised by  
\begin{eqnarray}
	U=\cos F(r,\theta)+i{\vec \tau}\cdot{\vec n}_R\sin F(r,\theta)
	\label{chiral}
\end{eqnarray}
with ${\vec n}_R = (\sin\Theta\cos n\varphi, \sin\Theta\sin n\varphi, \cos\Theta)$ 
and $\Theta =\Theta (r,\theta )$.
The integer $n$ corresponds to the winding number of the Skyrme fields and for $B=2$ we have $n=2$.

In terms of $F$ and $\Theta$, the Lagrangian takes the form 
\begin{eqnarray}
	{\cal L}_S={\cal L}_{S}^{(1)}+{\cal L}_{S}^{(2)} \label{}
\end{eqnarray}
where 
\begin{eqnarray*}
	{\cal L}_S^{(1)}& =& -\frac{f_\pi^2}{8}\frac{f}{m}\left[(\partial_{r}F)^2
	+\frac{1}{r^{2}}(\partial_{\theta}F)^2+\left\{(\partial_{r}\Theta)^2+\frac{1}{r^{2}}
	(\partial_{\theta}\Theta)^2\right\}\sin^{2}F \right. \\
	&& \left. + \frac{n^{2}}{r^2\sin^2 \theta}\frac{m}{l}\sin^{2}{\Theta}\sin^{2}F \right] \\
	{\cal L}_S^{(2)}&=&-\frac{1}{2a^{2}r^{2}}\left(\frac{f}{m}\right)^{2}
	\biggl[\left(\partial_{r}F\partial_{\theta}\Theta-\partial_{\theta}F
	\partial_{r}\Theta\right)^{2} \\
	&&+\frac{n^{2}}{\sin^{2}\theta}\frac{m}{l}\left\{(\partial_{r}F)^{2}
	+\frac{1}{r^{2}}(\partial_{\theta}F)^{2}\right\} \sin^{2}\Theta \\
	&&+ \frac{n^{2}}{\sin^{2}\theta}\frac{m}{l}
	\left\{(\partial_{r}\Theta)^{2}+\frac{1}{r^{2}}(\partial_{\theta}\Theta)^{2}
	\right\}\sin^{2}F\sin^{2}\Theta \biggr]\sin^{2}F \,\, . \label{}
\end{eqnarray*}
The baryon current in curved spacetime is obtained by 
taking the spacetime covariant derivative $\nabla_{\mu}$, 
\begin{eqnarray}
	b^{\mu}=\frac{1}{24\pi^{2}}\epsilon^{\mu\nu\rho\sigma}
	{\rm tr}(U^{-1}\nabla_{\nu}UU^{-1}\nabla_{\rho}U
	U^{-1}\nabla_{\sigma}U) .\label{}
\end{eqnarray}
The baryon number then is given by integrating $b^{0}$ 
over the hypersurface $t=0$, 
\begin{eqnarray}
	B&=& \int drd\theta d\varphi \sqrt{g^{(3)}}\;b^{0}  \nonumber \\
	&=& -\frac{1}{\pi}\int drd\theta\;(\partial_{r}F
	\partial_{\theta}\Theta-\partial_{\theta}F\partial_{r}\Theta)
	\sin\Theta(1-\cos 2F) \nonumber \\ 
	&=& \left. \frac{1}{2\pi}(2F-\sin 2F)\cos \Theta \right|_{F_{0},
	\Theta_{0}}^{F_{1},\Theta_{1}},\label{}
\end{eqnarray}
where $(F_{0},\Theta_{0})$ and $(F_{1},\Theta_{1})$ are the values at the inner 
and outer boundary respectively. 
In flat spacetime we have
\begin{eqnarray*}
	(F_{0},\Theta_{0})=(\pi,0)\;\; {\rm and}\;\; (F_{1},\Theta_{1})=(0,\pi) , \label{}
\end{eqnarray*}
which gives $B=2$.  
In the presence of a black hole, the integration 
should be performed from the horizon to infinity, which change 
the values of $F_{0}$ and allow the $B$ to take a fractional value of less than two. 
This situation can be interpreted as the black hole absorbing a skyrmion. 

The energy density is given by the time-time component of the 
stress-energy tensor  
\begin{eqnarray}
	-T_{0}^{0}&=& \frac{f_{\pi}^{2}}{8}\frac{f}{m}\left[(\partial_{r}F)^{2}
	+\frac{1}{r^{2}}(\partial_{\theta}F)^{2}+\left\{(\partial_{r}
	\Theta)^{2}+\frac{1}{r^{2}}(\partial_{\theta}\Theta)^{2}\right\}
	\sin^{2}F \right. \nonumber \\
	&& \left. +\frac{n^{2}}{r^{2}\sin^{2}\theta}\frac{m}{l}\sin^{2}F
	\sin^{2}\Theta\right] +\frac{1}{2a^{2}r^{2}}\frac{f^{2}}{m^{2}}
	\left[\frac{}{}(\partial_{[r}F\partial_{\theta]}\Theta)^{2} \right. \nonumber \\
	&& \left. +\frac{n^{2}}{\sin^{2}\theta}\frac{m}{l}
	\left\{(\partial_{r}F)^{2}+\frac{1}{r^{2}}(\partial_{\theta}F)^{2}
	\right\}\sin^{2}\Theta \right. \nonumber \\
	&& \left. +\frac{n^{2}}{\sin^{2}\theta}\frac{m}{l}
	\left\{(\partial_{r}\Theta)^{2}+\frac{1}{r^{2}}(\partial_{\theta}\Theta)^{2}
	\right\}\sin^{2}F\sin^{2}\Theta\right]\sin^{2}F.  
\end{eqnarray}

\section{Boundary Conditions}
Let us consider the boundary conditions for the chiral fields and metric functions. 

\subsection{Regular Solutions}
At the origin $r=0$, the metric functions are regular, which requires the boundary conditions 
\begin{eqnarray}
	\partial_{r}f(0,\theta)=\partial_{r}m(0,\theta)=\partial_{r}l(0,\theta)=0 . 
\end{eqnarray}
At infinity $r=\infty$, the metric is asymptotically flat. Thus we have  
\begin{eqnarray}
	f(\infty,\theta)=m(\infty,\theta)=l(\infty,\theta)=1.	
\end{eqnarray}
The boundary conditions for the Skyrme fields at $r=0, \infty$ are given by 
\begin{eqnarray}
      && F(0,\theta)=\pi, \;\; \partial_{r}\Theta(0,\theta)=0,  \\
	&& F(\infty,\theta)=0, \;\; \partial_{r}\Theta(\infty,\theta)=0. 
\end{eqnarray}
The axially symmetric condition leads to the boundary conditions on the axes $\theta =0, \pi/2$ as 
\begin{eqnarray}
 	&&\partial_{\theta}f(r,0)=\partial_{\theta}m(r,0)
	=\partial_{\theta}l(r,0)=0, \\
	&&\partial_{\theta}f(r,\frac{\pi}{2})=\partial_{\theta}m(r,\frac{\pi}{2})
	=\partial_{\theta}l(r,\frac{\pi}{2})=0,   \\
	&& \partial_{\theta}F(r,0)=0, \;\; \Theta(r,0)=0, \\
	&& \partial_{\theta}F(r,\frac{\pi}{2})=0, \;\; \Theta(r,\frac{\pi}{2})=\frac{\pi}{2}. \label{}	
\end{eqnarray}

  \subsection{Black Hole Solutions}
At the horizon $r=r_{h}$, the time-time component of the metric satisfies 
\begin{eqnarray}
	g_{tt}=-f(r_{h},\theta)=0 . \label{}
\end{eqnarray}
Regularity of the metric at the horizon requires  
\begin{eqnarray}
	m(r_{h},\theta)=l(r_{h},\theta)=0 . \label{}
\end{eqnarray}
The boundary conditions for $F(r,\theta)$ and $\Theta(r,\theta)$ at the horizon 
are obtained by expanding them at the horizon and inserting into the field equations 
which are derived from $\delta {\cal L}_{S}/\delta F=0$ and $\delta {\cal L}_{S}/\delta \Theta=0$ 
respectively. Thus we have  
\begin{eqnarray}
	\partial_{r}F(r_{h},\theta)=\partial_{r}\Theta(r_{h},\theta)=0.\label{}
\end{eqnarray} 
The condition that the spacetime is asymptotically flat requires
\begin{eqnarray}
	 f(\infty,\theta)=m(\infty,\theta)=l(\infty,\theta)=1 . \label{}
\end{eqnarray}
The boundary conditions for $F$ and $\Theta $ at infinity remain the same 
as in flat spacetime      
\begin{eqnarray}
	F(\infty,\theta)=0,\; \partial_{r}\Theta(\infty,\theta)=0 . \label{}
\end{eqnarray}
For the solution to be axially symmetric, we have 
\begin{eqnarray}
 	&&\partial_{\theta}f(r,0)
	=\partial_{\theta}m(r,0)
	=\partial_{\theta}l(r,0)=0 , \\
	&&\partial_{\theta}f\left(r,\frac{\pi}{2}\right)
	=\partial_{\theta}m\left(r,\frac{\pi}{2}\right)
	=\partial_{\theta}l\left(r,\frac{\pi}{2}\right)
	=0 . \label{}
\end{eqnarray}
Regularity on the axis and axisymmetry impose the boundary conditions on $F$ and $\Theta$ as  
\begin{eqnarray}
	&& \partial_{\theta}F(r,0)=\partial_{\theta}F\left(r,\frac{\pi}{2}\right)=0, \\
	&& \Theta(r,0)=0, \; \Theta\left(r,\frac{\pi}{2}\right)=\frac{\pi}{2} . \label{}
\end{eqnarray}
Under these boundary conditions, we shall solve the Einstein equations 
and the matter field equations numerically. 

\section{Numerical Results}

Let us introduce dimensionless coordinate and coupling constant   
\begin{eqnarray*}
	x=af_{\pi}r , \;\;	\alpha = \pi G f_{\pi}^{2} .\label{}
\end{eqnarray*}
Then the free parameters are the horizon $x_{h}$ and the coupling constant $\alpha$ 
for black hole solutions while for regular solutions the $\alpha$ is the only free parameter.  
We shall take $\alpha = 0$ as decoupling of gravity from the matter, effectively $G=0$.

\begin{figure}
\includegraphics{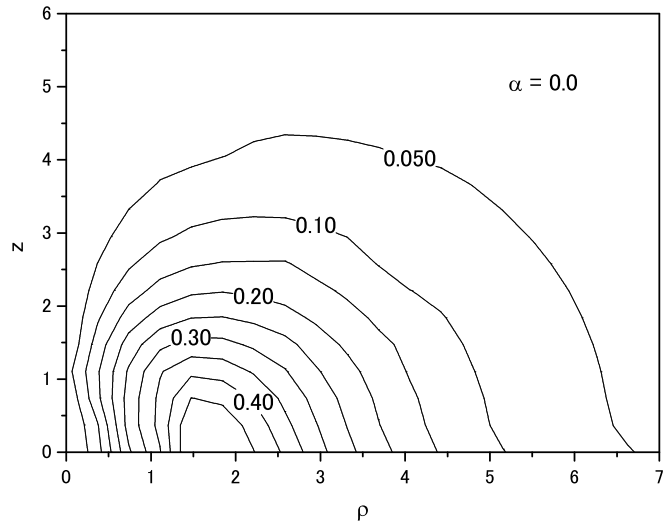}
\caption{\label{fig:reg_b0} Regular solution: The baryon density $b$ 
in cylindrical coordinates $\rho$ and $z$ with $\alpha =0.0$. } 
\end{figure}

In figs.~\ref{fig:reg_ed0}, \ref{fig:reg_ed3} are the energy densities per unit 
$r$ and $\theta $ $(\epsilon =- \sqrt{g^{(3)}}T^{0}_{0})$ of the regular solutions with 
$\alpha = 0.0$, $3.0$ respectively. As $\alpha $ becomes larger, the energy density gets  
concentrated in a smaller region. Thus the gravity influences for the matter to be more compact and denser. 
The shape remains torus irrespective of the value of $ \alpha $.  
We show the baryon density per unit $r$ and $\theta$ $(b=\sqrt{g^{(3)}}b^{0})$ 
in fig.~\ref{fig:reg_b0}. The shape is similar to the energy density, but its 
dependence on the $\alpha $ is rather small.   
Note that the energy density and the baryon density both vanish at $\rho =0$. 
This can be readily checked by inserting the Skyrme functions expanded around 
$\theta =0$ into $\epsilon $ and $b$ respectively.   
We found that there exists no regular solution for $\alpha \gtrsim 4.79$.  

\begin{figure}
\includegraphics{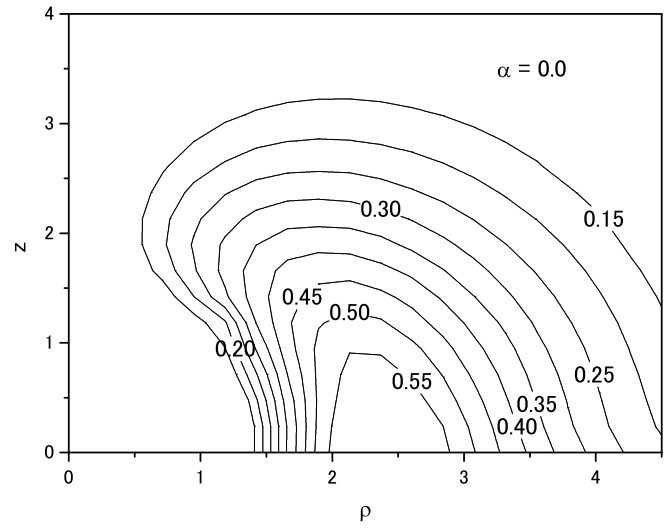}
\caption{\label{fig:b0} Black hole: The baryon density $b$ with $x_{h}=1.0, \alpha =0.0$. }
\end{figure}

In figs.~\ref{fig:ed0}, \ref{fig:ed15} are the energy densities of the black hole solutions 
with $\alpha = 0.0$, $1.5$ respectively. As $\alpha $ becomes larger, the energy density 
becomes smaller and sparse. This can be interpreted that the black hole absorbs more skyrmions  
for a larger coupling constant. The shape is slightly distorted in the background of the 
black hole so that one can see the spherically symmetric horizon in the center of the skyrmion. 
Fig.~\ref{fig:b0} is the baryon density around the black hole. As in the case of the regular 
solution, the dependence of the baryon density on the value of the coupling constant is small.  
It can be checked that the energy and baryon density 
vanish at $\rho =0 $ in the same manner as the regular solution. 
Inserting the metric functions as well as the Skyrme functions expanded around 
the horizon instead, one can also see that the energy and baryon density 
vanish at the horizon.    

\begin{figure}
\includegraphics{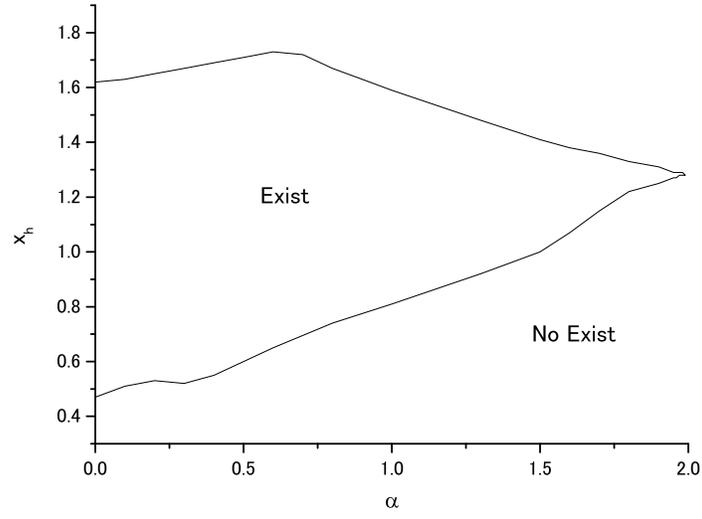}
\caption{\label{fig:domain} Black hole: The domain of existence of the 
solution. For $\alpha \gtrsim 2.0$, there exists no non-trivial 
solution.}
\end{figure}

The domain of existence of the black hole solution is shown in fig.~\ref{fig:domain}. 
There exist minimum and maximum value of $x_{h}$ and $\alpha$ beyond which no black hole solutions 
exist. Therefore the regular skyrmion solutions can not be recovered from the black hole solutions 
by taking the limit of $x_{h} \rightarrow 0$ unlike the case of $B=1$~\cite{Shiiki:bizon&chmaj}.
In fig.~\ref{fig:baryon_number} is the dependence of the baryon number on $x_{h}$. One can see that 
the baryon number decreases as the black hole grows in size.   
This figure confirms that the baryon number is no longer conserved due to the black hole 
absorbing the skyrmion. 

\section{Conclusions}

\begin{figure}
\includegraphics{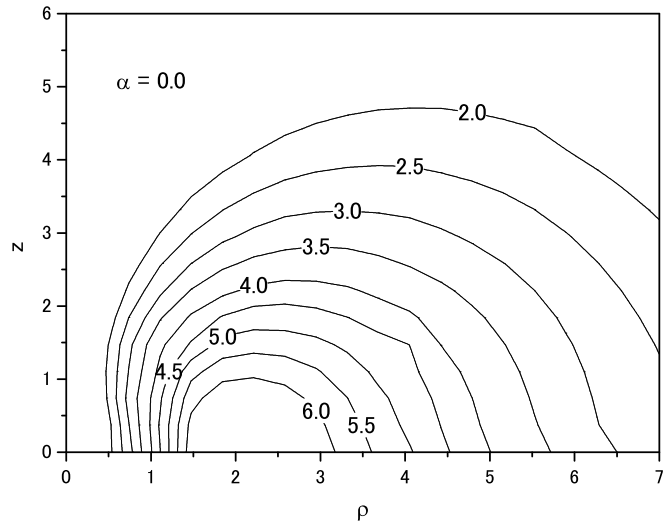}
\caption{\label{fig:reg_ed0} Regular solution: The energy density $\epsilon$ 
in cylindrical coordinates $\rho$ and $z$ with $\alpha = 0.0$. } 
\end{figure}

\begin{figure}
\includegraphics{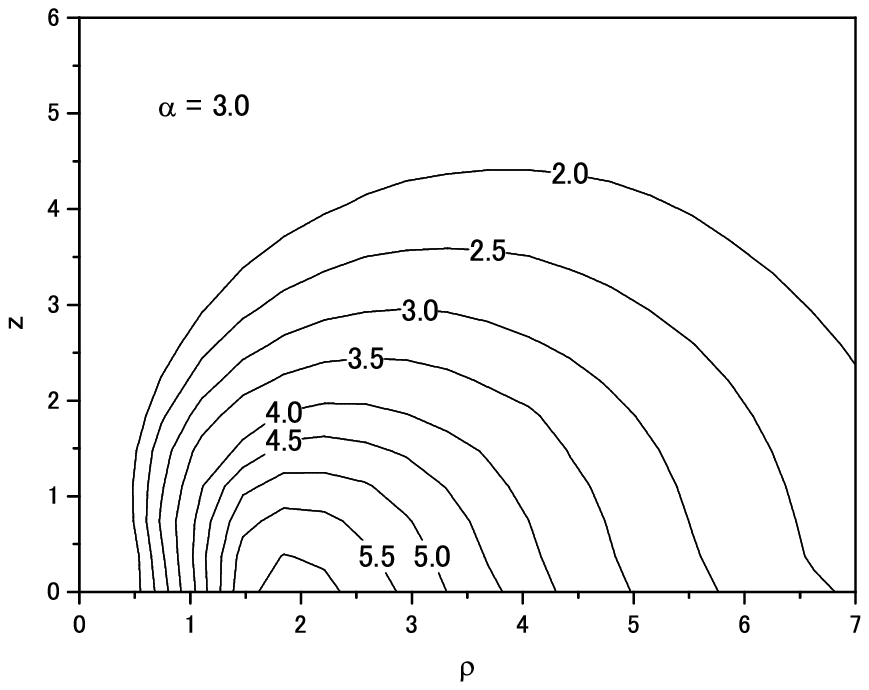}
\caption{\label{fig:reg_ed3} Regular solution: The energy density $\epsilon$ 
in cylindrical coordinates $\rho$ and $z$ with $\alpha =3.0$. } 
\end{figure}

\begin{figure}
\includegraphics{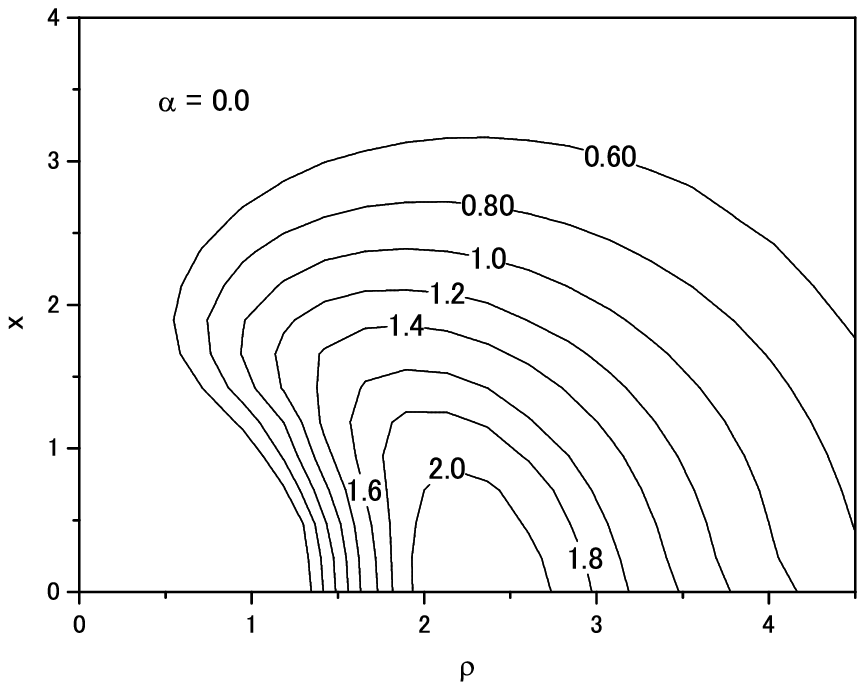}
\caption{\label{fig:ed0} Black hole: The energy density $\epsilon$ 
in cylindrical coordinates $\rho$ and $z$ with $x_{h}=1.0, \alpha = 0.0$. } 
\end{figure}

\begin{figure}
\includegraphics{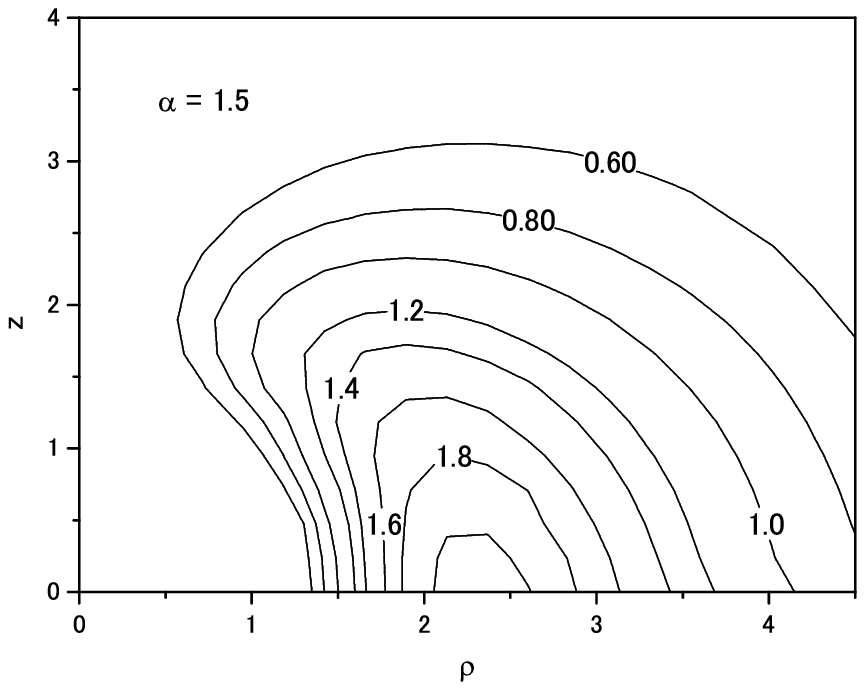}
\caption{\label{fig:ed15} Black hole: The energy density $\epsilon$ 
in cylindrical coordinates $\rho$ and $z$ with $x_{h}=1.0, \alpha = 1.5$. } 
\end{figure}

\begin{figure}
\includegraphics{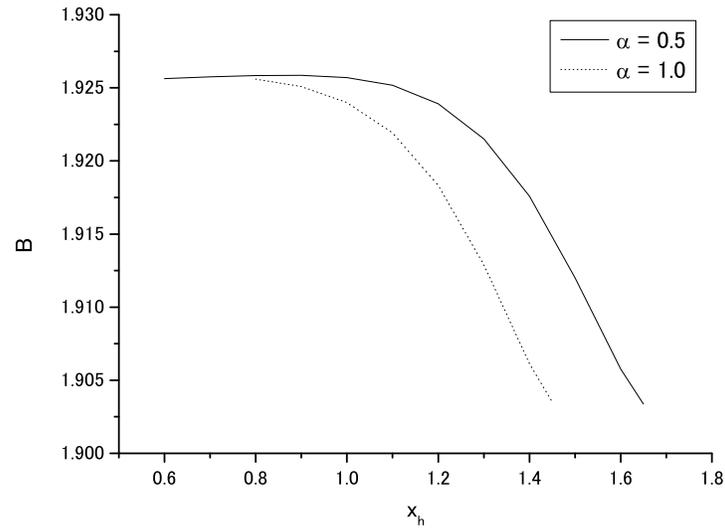}
\caption{\label{fig:baryon_number} Black hole: The dependence of the baryon number 
on the size of the horizon.}
\end{figure}

We have obtained static axially symmetric regular and black hole skyrmions numerically. 
As is shown in figs.~\ref{fig:reg_ed0}, \ref{fig:reg_ed3}, \ref{fig:ed0}, \ref{fig:ed15}, 
the energy density is toroidal in shape for the regular and black hole solutions. 
However, in the black hole case, the shape is distorted around the spherically 
symmetric horizon.    
The energy density depends on the coupling constant. 
For the regular solution, as the coupling constant becomes larger, 
the energy density distribute in a smaller region and become denser. 
On the other hand, for the black hole solution, the energy density becomes sparse 
due to the black hole absorbing more skyrmions.   
The baryon density shows a similar shape with the energy density and its dependence  
on the value of the coupling constant is rather small. However, one can see from 
fig.~\ref{fig:baryon_number} that it depends on the 
horizon size such that the baryon density decreases in increase of $x_{h}$. 
Since there are minimum values of $x_{h}$, the black hole solution 
can not recover a regular solution by taking the limit of $r_{h} \rightarrow 0$ 
unlike the $B=1$ spherically symmetric case. 
        
Obviously it is important to study the stability of our solutions. 
We expect that the stability analysis may be performed by applying the catastrophe 
theory for black holes with non-linear hair~\cite{Shiiki:maeda&tachizawa&torii&maki}.  

As our further works, it will be interesting to consider skyrmion black holes with 
$B \ge 3$ which have discrete symmetries. 
The inclusion of gauge fields may also be possible to study the interaction 
between a monopole black hole and a deuteron~\cite{Shiiki:moss&shiiki&winstanley}. 
Extending the model to higher dimensions will be also an exciting problem,  
leading to the study of deuteron black holes which may be observed in LHC 
in future~\cite{Shiiki:arkani-hamed&dimopoulos&dvali,Shiiki:antoniadis&arkani-hamed&dimopoulos,
Shiiki:banks&fischler}.


\end{document}